\documentclass[12pt]{article}
\usepackage{graphicx}

\def\overlay#1#2{\ifmmode%
\setbox0=\hbox{$#1$}%
\setbox1=\hbox to\wd0{\hss$#2$\hss}\else%
\setbox0=\hbox{#1}%
\setbox1=\hbox to\wd0{\hss#2\hss}\fi%
#1\hskip-\wd0\box1 }

\begin{document}

\hfill
{\vbox{
\hbox{UTEXAS-HEP-96-16}
\hbox{DOE-ER-40757-085}}}

\begin{center}
{\Large \bf Probing Non-standard Top Couplings Using spin-correlation}

\vspace{0.1in}

Kingman Cheung\footnote{Electronic mail: {\tt cheung@utpapa.ph.utexas.edu}}

{\it  Center for Particle Physics, University of 
Texas, Austin, Texas 78712}
\end{center}

\thispagestyle{empty}

\begin{abstract}
Top spin correlation has been shown to be nontrivial in hadronic 
top-pair production, and can be realized by the  asymmetries of
the decay products of the top quark and antiquark.  We show in 
this work that the top spin correlation is a sensitive probe to the 
anomalous top-quark couplings beyond the standard model.  Specifically, 
we look at the anomalous chromomagnetic and chromoelectric dipole 
moments, as well as a right-handed component in the weak decay of the
top quark.  We found a few measurable asymmetries formed by the decay 
products of the top-quark pair that vary in accord with the spin correlation.
\end{abstract}

\section{Introduction}
The top quark is very different from the other five quarks because it has
a mass \cite{topmass} close to the electroweak symmetry breaking scale.
The top quark can therefore provide useful avenues to probe the physics
beyond the standard model (SM) through its decay, direct production, 
scattering, and the associated production with other particles.
Since the structure or even the symmetry of the correct high energy theory
is not known,   the effective Lagrangian approach can be used 
 to study  low energy phenomena. 
Deviations from the SM can be studied systematically by including higher 
dimensional operators, which are made up of the SM fields, into the interaction
Lagrangian.  Such higher dimensional operators are suppressed by powers of 
the scale $\Lambda$ of the new physics. 

In this paper, we study some anomalous couplings of the top quark, which 
often appear in extensions of the SM.  We shall limit ourselves to as
low dimension as possible because the effects of the high dimensional 
operators are very much suppressed at low energies. 
Here, we concern only the dim-4 and dim-5 
operators.  For dimension 4 we include a
right-handed component into the weak decay of the top quark.  Although the 
$V-A$ structure in light quarks are more or less established, the $V-A$
structure of the top quark should be confirmed.  The presence of a right-handed
component, even small, will signify some new physics beyond the SM.
For dimension 5 we concentrate on the anomalous chromomagnetic and 
chromoelectric dipole moments of the top quark.  They are particularly 
interesting because they are directly involved in the subprocesses
$gg,q\bar q\to t\bar t$ of top-pair production, and they are often
nonzero on loop-level or even tree level in many extensions of the SM,
e.g., the multi-Higgs doublet model \cite{weinberg}.
Furthermore,  a nonzero value for the 
chromoelectric dipole moment is a clean signal of CP violation.  The effects
of these anomalous couplings have been studied in $t\bar t$ production
\cite{cheung,rizzo,haberl}, $b\bar b$ production \cite{rizzo}, inclusive
jet production \cite{dennis}, and prompt photon production \cite{photon}.

In this work, we study the effects of the above mentioned anomalous
couplings on the spin correlation in top-pair production.
Recently, a few studies \cite{tim,greg,darwin}
showed that top-pair production is highly 
spin-correlated, which means that the top quark and antiquark have 
preferential spin polarizations.  
A theoretically interesting variable to quantify the spin correlation is 
\cite{tim}
\begin{equation}
\label{c}
C = \frac{\sigma(t_L \bar t_L + t_R \bar t_R) - 
          \sigma(t_L \bar t_R + t_R \bar t_L) }
         {\sigma(t_L \bar t_L + t_R \bar t_R) +
          \sigma(t_L \bar t_R + t_R \bar t_L) } \;,
\end{equation}
where the subscripts $L$ and $R$ denote the helicities of the top quark and
antiquark.  At the Tevatron, the dominant subprocess
of top-pair production is $q\bar q  \to t\bar t$ and $C\approx -0.4$, while 
at the LHC energy $gg\to t\bar t$ will dominate and $C\approx +0.3$ \cite{tim}.
The measurement of $C$ can then serve as an indirect probe to the 
underlying production mechanism.
Although this variable is a true measure of the spin correlation, it is 
not directly measured experimentally.  Fortunately, the top quark is
so heavy that it decays before it hadronizes, and, therefore, almost
all the spin information is retained in the decay products of the top quark. 
The heavy top quark will decay preferentially into a 
longitudinally polarized $W^+$ boson and a left-handed $b$ quark.  Because of
the conservation of angular momentum 
the $W^+$ boson prefers to go in the same direction as the top polarization
in the rest frame of the top quark.  The antilepton $\ell^+$ decaying from
the $W^+$ boson also goes in the direction of the $W^+$ boson, i.e.,
in the same direction as the spin polarization of the top quark.  
On the other hand, 
the lepton decaying from the top antiquark prefers to go in the opposite 
direction as the spin polarization.  Thus, by discriminating the directions of
the lepton and antilepton we can select particular polarizations of the top
quark and antiquark.  
Since the top-quark pair is spin-correlated, the asymmetries formed by
the lepton and antilepton could be nontrivial.  We shall introduce a few
asymmetries in the next section.

The organization is as follows.  In Sec. II we introduce the asymmetries and 
calculate them in the SM.  In Sec. III we study the effects of the anomalous 
chromomagnetic and chromoelectric dipole moments of the top quark on the 
variable $C$ and the asymmetries, and in Sec. IV we study the effects
of the right-handed component in the weak decay of the top quark.  We
then discuss and conclude in Sec.~V.
In this work, we  concentrate on the Run II of the Tevatron ($\sqrt{s}=
2$ TeV) with a yearly luminosity of order $2\;{\rm fb}^{-1}$ \cite{tev2000}.  
The top quark mass is chosen to be 175 GeV \cite{topmass}.  We use the parton 
distributions of CTEQ3L \cite{cteq}.

\section{Spin Correlation and Asymmetries within the SM}

The dominant subprocesses for the hadronic production of top quark are
\begin{eqnarray}
q \bar q \to t \bar t \nonumber \\
gg \to t \bar t \;.\nonumber 
\end{eqnarray}
The $q\bar q$-initiated subprocess dominates at the Tevatron energies (
$\sqrt{s}=1.8$ TeV for Run I and 2 TeV for Run II), while $gg$-initiated one
dominates at the LHC, because of the increasing gluon luminosity with increases
in energy.   At the Tevatron energies, since most of the times the top pair is 
produced near threshold and in the dominant subprocess 
$q\bar q \to t\bar t$  the top pair is produced via a $J=1$ $s$-channel gluon,
most of the $t\bar t$ pairs are in a $^3S_1$ state.  
By counting the spin eigenstates
the ratio of the top-quark pair in the same helicities to the top-quark 
 pair in opposite
helicities is 1:2, so the value of $C$ is $-\frac{1}{3}$.  Far above the
threshold, helicity conservation requires $t$ and $\bar t$ in opposite 
helicities, so $C=-1$.  We found that at the Tevatron ($\sqrt{s}=1.8-2.0$ TeV)
$C\approx -0.4$, which was first given in Ref.~\cite{tim}.  On the other hand,
$gg\to t\bar t$ dominates at the LHC.  Near threshold the $t\bar t$ pair is 
produced in a $^1S_0$ state, so the $t$ and $\bar t$ have the same helicities,
with $C=+1$.  Hence, there is a dramatic change of $C$ from negative 
to positive when energy increases from 1.8 TeV to 14 TeV.  

As mentioned above that although $C$ is not directly measured, it can be
revealed in the decay products of the top-quark pair.
The best analyzer of the top spin polarization is the semileptonic decay,
because in other decay modes it is difficult to 
distinguish the up-type or down-type quark in the decay
of the $W$ boson, and leptons are much easier to be detected.  
We will study three different asymmetries, two of 
which have been shown to be very effective \cite{tim,darwin}.   
All three of them involve semileptonic decays of the top
quark and antiquark.  The first one, denoted by $A$, is \cite{tim}
\begin{equation}
A = \frac{\sigma(z_1 z_2 >0) - \sigma(z_1 z_2 <0 )}
         {\sigma(z_1 z_2 >0) + \sigma(z_1 z_2 <0 )}\;,
\end{equation}
where $z_i = \cos\theta_i\; (i=1,2)$ is the cosine of the angle of the lepton
in the rest frame of the parent top quark 
with respect to the momentum of the top
in the $t\bar t$ center-of-mass frame.  The second one, denoted by $B$, 
is 
\begin{equation}
B = \frac{\sigma(p_1 p_2 >0) - \sigma(p_1 p_2 <0 )}
         {\sigma(p_1 p_2 >0) + \sigma(p_1 p_2 <0 )}\;,
\end{equation}
where $p_1$ and $p_2$ are the $z$-component of the 3-momenta of the lepton
and antilepton in the $t\bar t$ center-of-mass frame.  
The positive $z$ direction is defined as the direction of incoming proton
beam.   In other words, this
asymmetry $B$ counts the number of events that the lepton and antilepton
go to the same or opposite side of the plane, which is normal to the beam
direction,  in the $t\bar t$ center-of-mass frame.   The third one, which 
we denote by $B'$ \footnote{$B'$ is exactly the same as $A_4$ described in 
Ref.~\cite{darwin}, in which $A_4$ is the best 
out of the four asymmetries considered.}, 
is given by \cite{darwin}
\begin{equation}
B' = \frac{\sigma(p'_1 p'_2 >0) - \sigma(p'_1 p'_2 <0 )}
         {\sigma(p'_1 p'_2 >0) + \sigma(p'_1 p'_2 <0 )}\;,
\end{equation}
where $p'_1$ ($p'_2$) is the $z$-component of the 3-momentum of the antilepton
(lepton) in the rest frame of the top quark (top antiquark).  
The only difference between $B$ and $B'$ is that the 3-momenta of the lepton
and antilepton are in different frames.

To calculate these asymmetries we have to put in the decay matrix elements of 
the top quark and antiquark 
with full spin correlation.  We use the helicity amplitude method
\cite{alan} to calculate the spin-polarized cross sections, and the asymmetries
$C$, $A$, $B$, and $B'$.  
With this helicity amplitude method the subsequent decays
of the top quark and the $W$ boson can be included straight-forwardly.  The 
necessary formulas for top-pair production and the decays are given in
Ref.~\cite{cheung}.  We found that the asymmetry $A$ is about $+0.1$ in 
accord with the results of Ref.~\cite{tim}, 
the asymmetry $B$ is about $-0.3$, and
$B'$ is about $-0.2$,  which also agrees with  the results of 
Ref.~\cite{darwin}.
Although the asymmetries $B$ and $B'$ are larger than $A$ numerically, 
$A$ is more sensitive to the
anomalous chromomagnetic dipole moment that we are going to discuss in the
next section.  With the asymmetries $C$, $A$, $B$, and $B'$ 
we can now study the
new physics associated with top-gluon vertex and the weak decay vertex 
of the top quark.  They are, respectively, studied in the next two sections.

\section{Chromomagnetic and Chromoelectric Dipole Moments}

The effective Lagrangian for the interactions between a top quark and a gluon
that include the chromomagnetic (CMDM) and chromoelectric (CEDM) form 
factors is 
\begin{equation}
\label{eff}
{\cal L} = g_s \bar t T^a \left[ -\gamma^\mu G_\mu^a +
\frac{\kappa}{4m_t} \sigma^{\mu\nu} G_{\mu\nu}^a - 
\frac{i \tilde{\kappa}}{4m_t} \sigma^{\mu\nu} \gamma^5 G_{\mu\nu}^a 
\right ] t \;,
\end{equation}
where $\kappa/2m_t$ ($\tilde{\kappa}/2m_t$) is the CMDM (CEDM) of the top 
quark.  The Feynman rules for the interactions of the top and gluon can be
written down:
\begin{equation}
\label{ttg}
{\cal L}_{t_i t_j g} = -g_s \bar t_j T^a_{ji} \left[ \gamma^\mu +
\frac{i}{2m_t} \sigma^{\mu\nu} p_\nu (\kappa - i \tilde{\kappa} \gamma^5) 
\right ] t_i\; G_\mu^a \;,
\end{equation}
where $t_i (t_j)$ is the incoming (outgoing) top and $p_\nu$ is the 4-momentum
of the outgoing gluon.  The Lagrangian in Eq.~(\ref{eff}) also induces a 
$ttgg$ interaction given by 
\begin{equation}
\label{ttgg}
{\cal L}_{t_i t_jgg} = \frac{ig_s^2}{4m_t} \bar t_j (T^b T^c - T^c T^b)_{ji}
\sigma^{\mu\nu} ( \kappa -i \tilde{\kappa} \gamma^5) t_i G_\mu^b G_\nu^c \;,
\end{equation}
which is absent in the SM.  

\begin{figure}[tbh]
\leavevmode
\begin{center}
\includegraphics[height=3.5in]{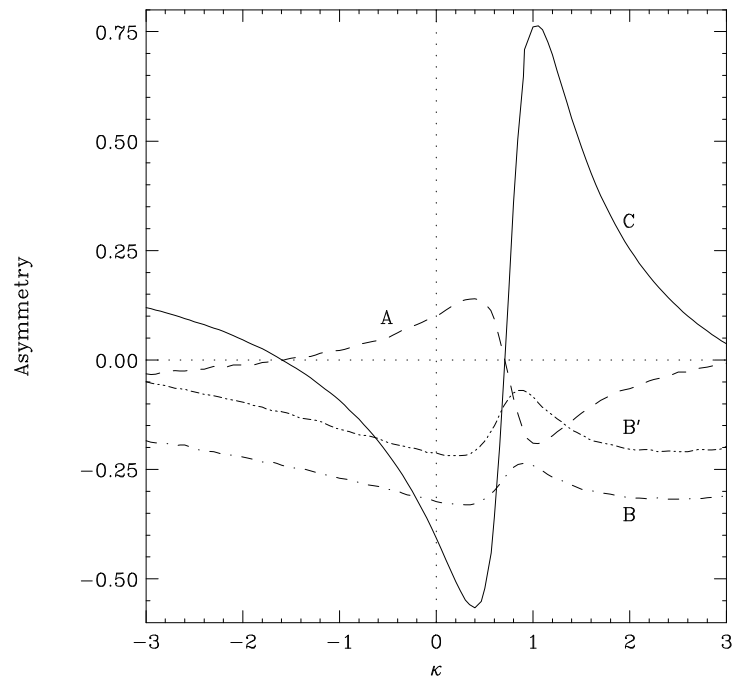}
\end{center}
\caption{Variations of the spin correlation $C$, and the asymmetries 
$A$, $B$, and $B'$ versus  $\kappa$.}
\label{fig1}
\end{figure}

\begin{figure}[th]
\leavevmode
\begin{center}
\includegraphics[height=3.5in]{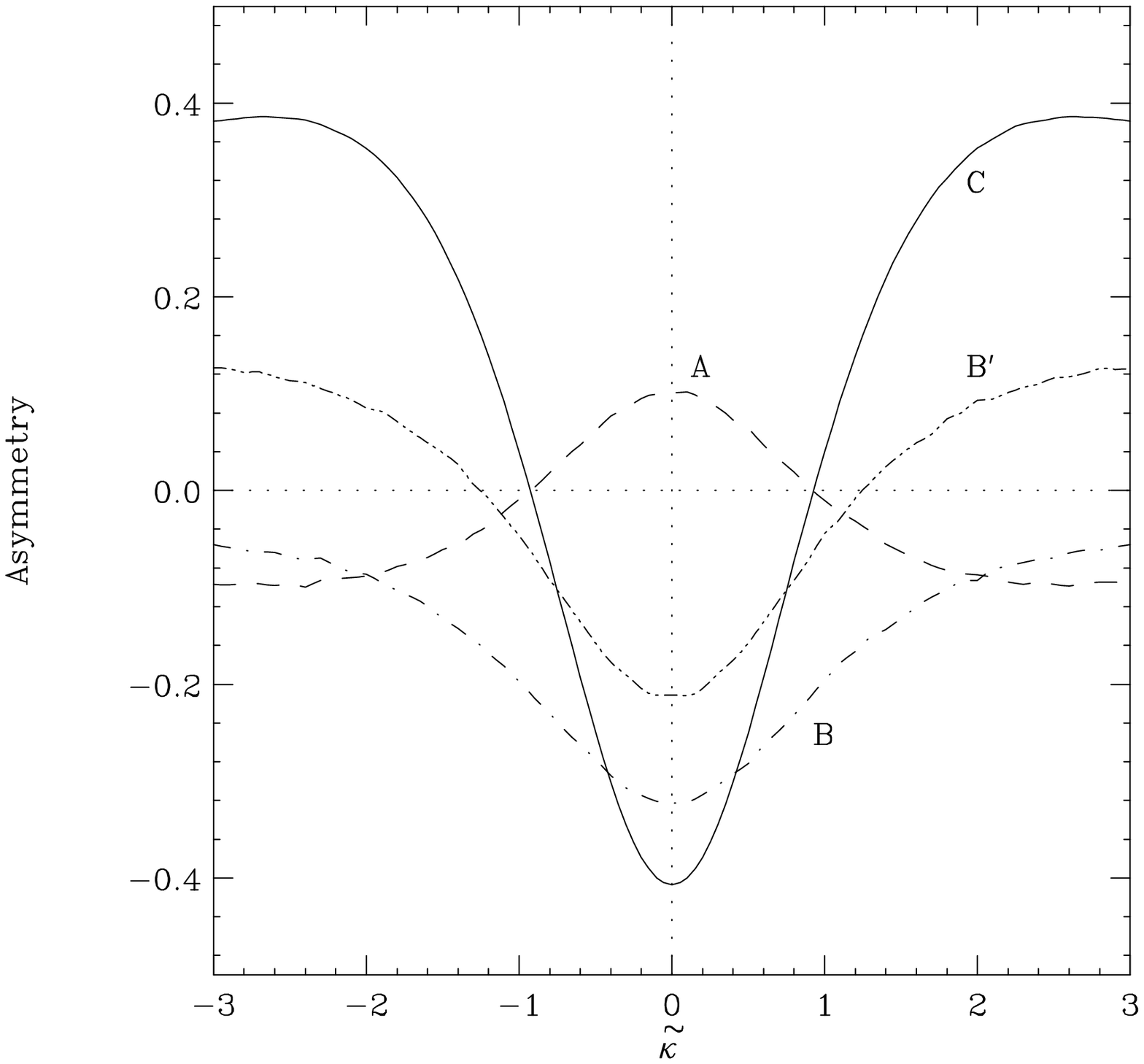}
\end{center}
\caption{Variations of the spin correlation $C$, and the asymmetries 
$A$, $B$, and $B'$ versus  $\tilde\kappa$.}
\label{fig2}
\end{figure}

The unpolarized $t\bar t$ cross sections including the CMDM and CEDM couplings
were calculated in Ref.~\cite{cheung,rizzo,haberl}.  Since we are interested 
in polarized cross sections here, we use the helicity amplitude method 
\cite{alan} to calculate the polarized cross sections including the CMDM
and CEDM couplings.
Once we obtained the polarized cross sections for $t_L\bar t_L$,
$t_R\bar t_R$, $t_L\bar t_R$, and $t_R\bar t_L$, we can then calculate the
ideal spin-correlation variable $C$ of Eq.~(\ref{c}).
We show the results of $C$  versus $\kappa$ and $\tilde \kappa$ by 
the solid curves in Fig.~\ref{fig1} and Fig.~\ref{fig2}, respectively.  
As expected the behavior of the curve for $\tilde \kappa$ is symmetric 
about the $y$-axis because the cross sections contain only even powers of
$\tilde \kappa$, while the curve for $\kappa$ is not symmetric about $y$-axis
because the cross sections do depend on odd powers of $\kappa$.
Since $C$ is so sensitive to the anomalous dipole moments, if one were
able to measure $C$ directly and accurately enough, we could measure
the CMDM and CEDM of the top quark easily.  However, this is not so.
But it is not impossible to have indirect measurements of $C$ by means of the
asymmetries $A$, $B$, and $B'$ that we introduced before.  
Similar to how we obtained the SM results, we decay the top quark and 
antiquark semileptonically with full spin correlation.
The variations
of $A$, $B$, and $B'$ with $\kappa$ and $\tilde \kappa$ are also superimposed 
with $C$ onto Fig.~\ref{fig1} and Fig.~\ref{fig2}.  From these two figures
 we can see
that the rise and fall of the curves $A$, $B$, and $B'$ are in 
accord with those of $C$.  
It is then clear that the asymmetries $A$, $B$, and $B'$ are, to a
great extent, true representations of the spin-correlation $C$. 
Especially, $A$ is directly related to $C$ by $A=-C/4$ \cite{tim}.
We also see that the asymmetry $A$ is more sensitive than the asymmetries 
$B$ and $B'$ to changes in $\kappa$, while all three  are more or less equally
 sensitive to changes in $\tilde \kappa$.  Nevertheless, 
the asymmetry $B$ has a larger numerical value than $A$ and $B'$ at
the SM point  ($\kappa=\tilde\kappa=0$), which 
implies that it statistically needs fewer dilepton events to see
 the spin correlation.

\section{Right-handed Weak Decay}

\begin{figure}[th]
\leavevmode
\begin{center}
\includegraphics[height=3.5in]{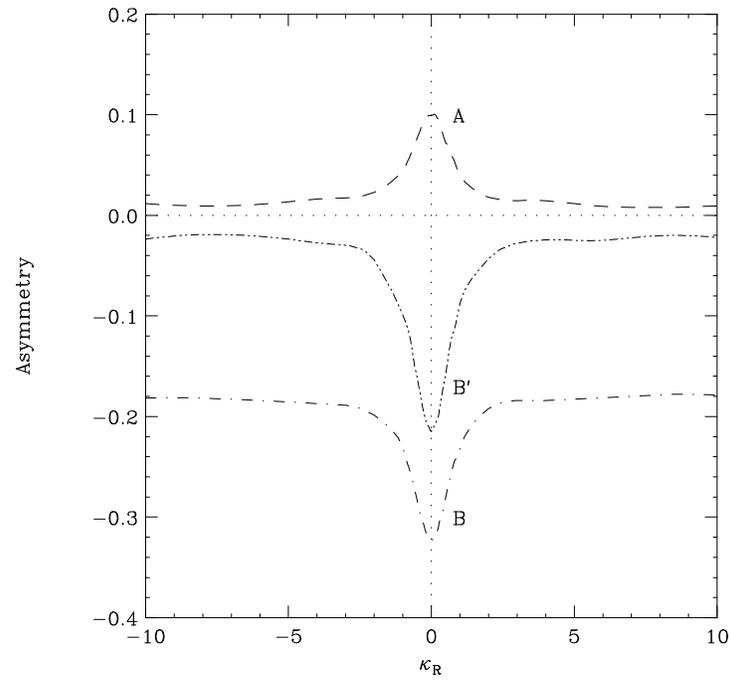}
\end{center}
\caption{Variations of the asymmetries $A$, $B$, and $B'$ versus $\kappa_R$.}
\label{fig3}
\end{figure}

In dimension 4 the general charged-current couplings of the top quark is 
given by \cite{yuan}
\begin{equation}
{\cal L} = -\frac{g}{\sqrt{2}}\, \bar b \gamma^\mu \biggr[ \frac{1}{2}
(1-\gamma^5) (1+\kappa_L) + \frac{1}{2}(1+\gamma^5)\kappa_R \biggr ] t W_\mu^-
\; + \; {\rm h.c.}
\end{equation}
where the parameters $\kappa_L$ and $\kappa_R$ are used to denote the strength
of the additional left-handed and right-handed couplings, respectively.
In the SM, $\kappa_L=\kappa_R=0$.
Since the right-handed weak decay interaction only affects the decays of the 
top quark and antiquark, it has no effects on the variable $C$, as $C$ only 
depends on the top-gluon couplings.  
But it will affect the asymmetries $A$, $B$, and $B'$. 
The non-standard top decay 
can be implemented using the helicity amplitude method \cite{alan}, replacing
the spinors of the top quark and top antiquark by
\begin{eqnarray}
\bar u(t) &\to & -\frac{g^2}{8} \, \frac{1}{W^2 -m_W^2 + i\Gamma_W m_W} \,
\frac{1}{t^2-m_t^2 + i \Gamma_t m_t} \, 
\bar u(\nu)\gamma_\mu (1-\gamma_5)v(\ell^+)  \nonumber  \\
&& \times \;\;\; \bar u(b)\gamma^\mu \, \biggr (
1+\kappa_L + \kappa_R +\gamma^5(\kappa_R -1 -\kappa_L) \biggr)\, 
(\overlay{/}{t} + m_t )\; \\
v(\bar t) &\to & -\frac{g^2}{8} \, \frac{1}{W^2 -m_W^2 + i\Gamma_W m_W} \,
\frac{1}{{\bar t}^2-m_t^2 + i \Gamma_t m_t} \; 
\bar u(\ell^-)\gamma_\mu (1-\gamma_5)v(\bar \nu) \nonumber \\
&& \times \;\;\; (- \overlay{/}{\bar t} + m_t )\, \gamma^\mu 
\biggr ( 1+\kappa_L + \kappa_R +\gamma^5(\kappa_R -1 -\kappa_L) \biggr)\, 
v(\bar b) \;  
\end{eqnarray}
where the 4-momenta of the particles 
are labeled by the particle symbols.  We have used the 
narrow width approximation to handle the top and $W$ boson propagators, and 
we used the SM values for the width of the top quark and the $W$ boson.
Since we are dealing with the asymmetries only (not the total cross sections),
we do not need to include the non-standard interactions in calculating the
top width $\Gamma_t$.
We have verified that changing $\kappa_L$ alone while keeping $\kappa_R=0$ 
would not change the asymmetries because the weak decay is still $V-A$.
Therefore, we only show the results versus $\kappa_R$ while keeping 
$\kappa_L=0$.   In Fig.~\ref{fig3}, we show the asymmetries $A$, $B$, 
and $B'$ versus $\kappa_R$ from $-10$ to 10.  We can see that all three
asymmetries decrease for non-zero $\kappa_R$,
and are symmetric about $\kappa_R=0$.  Even at $\kappa_R=1$ (vector
coupling only)  and at $\kappa_R=-1$ (axial-vector coupling only) the
asymmetries are non-zero, because non-zero spin correlation is still being
fed down from top-pair production ($C\approx -0.4$ in the SM).
As $\kappa_R$ increases further away from zero, the asymmetries $A\to +0.01$,
$B\to -0.18$, and $B' \to -0.02$.  It would also be interesting
to check the case of an entirely right-handed coupling by putting 
$\kappa_L=-1$ and $\kappa_R=1$, at which $A$, $B$, and $B'$ are verified 
to be the same as the asymptotic values of the curves in 
Fig.~\ref{fig3}.
The reason why $A$ and $B'$ are not quite zero yet is that the $t\bar t$ 
pairs being produced are still spin-correlated, i.e., $C\approx -0.4 \not =
0$.  Nevertheless, the right-handed weak decay of the top quark and 
antiquark causes the asymmetries very difficult to be detected.
We have also verified that at the point where $C=0$, i.e., the spins of 
the top pair are uncorrelated, the values of $A$ and $B'$ are in fact zero
for $\kappa_L=-1=-\kappa_R$.

\section{Discussions}

\noindent
{\bf (a)}
So far, the calculation of the asymmetries are without experimental acceptance
cuts, efficiencies of reconstructing the 4-momenta of the top quark and
antiquark, or the smearing of momenta due to the detector.  
Typical acceptance cuts on the observed leptons, $b$ quarks, and missing 
transverse energy are
\begin{equation}
\label{cut}
p_T(b,\ell) > 15 \; {\rm GeV} \;, \qquad  |y(b,\ell)|< 2\;, \qquad 
\not{p}_T > 25 \;{\rm GeV} \;,
\end{equation}
which are also needed for eliminating backgrounds and for reconstructing 
the top quark and antiquark. 
We found that the cuts reduce the asymmetries by about 15--20\% near 
the SM point ($\kappa=\tilde\kappa= \kappa_R=0$).
The effects of smearing and reconstruction of the top quark and antiquark
rest frame have been studied in details in Ref.~\cite{darwin}, so we do not
repeat here.  In general, the asymmetries near the SM point
are reduced by another 10--15\%.  However, since it is only 
based on a parton-level Monte Carlo, a full Monte Carlo is needed to study
the true effects.

\vskip0.2in
\noindent
{\bf (b)}
The SM values for the asymmetries $A$, $B$, and $B'$ are $+0.1$, $-0.3$, and
$-0.2$, respectively.  It would be important to check if these asymmetries
can be observed above experimental uncertainties.
Since we do not have a full Monte Carlo simulation, we only take into account
the statistical error.  The SM cross section for $t\bar t$ production
with the dilepton decay mode at $\sqrt{s}=2$ TeV under the cuts in
Eq.~(\ref{cut}) is about 0.14 pb, which gives about 280 dilepton events for 
a $2\;{\rm fb}^{-1}$ luminosity in the Run II of the Tevatron \cite{tev2000}.
The statistical error is then $\sqrt{280}/280\approx 0.06=6\%$.  
Therefore, the values for $A=+0.1$, $B=-0.3$, and $B'=-0.2$ should be
clean to be observed, even after taking into account other systematical 
uncertainties, especially, $B'$ shows a possible $3\sigma$ effect from an
uncorrelated $t\bar t$ sample.   Thus, the spin correlation in $t\bar t$ 
production can be tested cleanly in the Run II of the Tevatron.

\vskip0.2in
\noindent
{\bf (c)}
In the following, we are going to estimate the sensitivities to or the bounds
on $\kappa,\tilde\kappa$, and $\kappa_R$ using the various spin-correlation
asymmetries if assuming the SM is correct.  
To estimate the $1\sigma$ sensitivity to $\kappa,\tilde\kappa$, and 
$\kappa_R$, we assume the SM values for $A$, $B$, and $B'$ are correct and 
put a $\pm 0.06$ onto them to get the corresponding ranges for
$\kappa,\tilde\kappa$, and $\kappa_R$.  The bounds are then given by
\begin{equation}
\begin{array}{rcl}
-0.7 < & \kappa & < +0.6 \\
-0.5 < & \tilde \kappa & < +0.5 \\
-0.5 < & \kappa_R & < +0.5 \;,
\end{array}
\end{equation}
where we have combined, for each of the $\kappa$'s, the three ranges given
by $A$, $B$, and $B'$.
Although these estimates are rather crude, they do give a feeling of how 
well the limits can be obtained using the spin-correlation.
In reality, the limits should be weaker than the above because there are
also systematical errors which have to be taken into account.
Nevertheless, 
the limits obtained above are comparable to those using total cross sections.
Moreover, using spin-correlation is better than using the total cross section
in controlling the uncertainties coming from higher order corrections,
parton distribution functions, and the strong coupling constant. 

Higher luminosities, e.g., an integrated $10\;{\rm fb}^{-1}$ in the 
stretched run of the Run II or even $100\;{\rm fb}^{-1}$ in TeV33 plan 
\cite{tev2000} will certainly reduce the statistical error by 
the square root of the increase in luminosity.  Using a $
10\; (100)\;{\rm fb}^{-1}$
luminosity compared to a $2\;{\rm fb}^{-1}$ luminosity reduces the statistical
error by a factor of $\sqrt{5} \;(\sqrt{50})$.  Thus, the limits can also be
improved substantially.
There is also a possibility of using other decay modes of the top quark
and antiquark, which can result in a larger number of events, thus
reducing the statistical error.  However, it is very difficult experimentally
to distinguish the quark and antiquark in the $W$ boson decay, which then
reduces the asymmetries significantly \cite{tim}.

\vskip0.2in
\noindent
{\bf (d)}
Other facilities to study the top quark are the $e^+e^-$ colliders
at 0.5 TeV and the LHC.  In $e^+e^-$ collisions, since the production of
$t\bar t$ pair is via $s$-channel exchanges of $\gamma$ and $Z$, the 
production rate of $t\bar t$ actually decreases with increase in energy
when the energy is well above the threshold.  Therefore, it would not be
advantageous to study $t\bar t$ production in very high energy $e^+e^-$
colliders. 
The $t\bar t$ pair produced in $e^+e^-$ collisions will also be 
spin-correlated, and the studies in this paper can be applied.  We can 
further study the variation versus the center-of-mass energy of the 
collisions. 

The LHC will be a copious source of $t\bar t$ pair, of order $10^6-10^7$.
There should be large number of dilepton events to measure the 
spin-correlation asymmetries down to one percent accuracy.  Furthermore,
the dominant production process changes to $gg\to t\bar t$, thus also changing
the spin-correlation substantially, as discussed in the Introduction.

In conclusions, we have studied the spin correlation of $t\bar t$
production and the asymmetries formed by the decay products of the top quark
and antiquark at the Fermilab Tevatron.  We also studied the effects
of anomalous chromomagnetic and chromoelectric dipole moments of the top
quark, and a right-handed component in the weak decay of the top quark on
the spin correlation.  We also estimated the limits on $\kappa$, 
$\tilde\kappa$, and $\kappa_R$ that can be statistically obtained in the 
Run II of the Tevatron if assuming the SM is correct.

We acknowledge the support from the U.S. Department of Energy under grant no.
DE-FG03-93ER40757.

% --------------------------------------

%%%%%%%%%%%%%%%%%%%%%%%%%%%%%%%%%%%%%%%%%%%%%%%%%%%%%%
\vspace{2in}

\end{document}